\documentclass[showpacs,preprintnumbers,prd,showkeys,aps]{revtex4}

\usepackage{braket}
\usepackage{eurosym}
\usepackage{amssymb}
\usepackage{amsfonts}
\usepackage{amsmath}
\usepackage{graphicx}
\usepackage{calrsfs}
\usepackage{color}
\usepackage[usenames,dvipsnames,svgnames,table]{xcolor}
\usepackage[colorlinks=true,
            linkcolor=red,
            urlcolor=red,
            citecolor=blue]{hyperref}

\begin{document}

\title{Light Deflection with Torsion Effects Caused by a Spinning Cosmic String}
\author{Kimet Jusufi}
\email{kimet.jusufi@unite.edu.mk}
\affiliation{Physics Department, State University of Tetovo, Ilinden Street nn, 1200,
Macedonia}
\date{\today }

\begin{abstract}
Using a new geometrical method introduced by Werner, we find the deflection angle in the weak limit approximation by a spinning cosmic string in the context of the Einstein-Cartan (EC) theory of gravity. We begin by adopting the String-Randers optical metric, then we apply the Gauss-Bonnet theorem to the optical geometry and derive the leading terms of the deflection angle in the equatorial plane. Calculations shows that light deflection is affected by the intrinsic spin of the cosmic string and torsion.
\end{abstract}
\pacs{11.27.+d, 95.30.Sf, 04.50.+h, 02.40.Hw}
\keywords{Light deflection, Cosmic string, Torsion, Finsler geometry, Gauss-Bonnet theorem}
\maketitle
\section{Introduction}

Gravitational bending of light by massive object is a well known phenomena which led to the first experimental evidence of the general theory of relativity \cite{eddington}. Remarkably, even today, the deflection of light continues to be one of the most important tools used in modern astrophysics and cosmology. This phenomena has been studied in details in various astrophysical aspects, both in the weak limit \cite{weak1,weak2} and strong limit approximation \cite{strong1,strong2,strong4}. In the last few years, there has been a growing interest in studying weak as well as strong field, along this line of research many papers have been written, including, the naked singularities and relativistic images of Schwarzschild black hole lensing, role of the scalar field in gravitational lensing, \cite{virbhadra1,virbhadra2,virbhadra3}, wormholes \cite{nandi}, to test the cosmic censorship hypothesis \cite{deandrea}, gravitational lensing from charged black holes in the weak field limit \cite{sereno1}, Kerr black hole lensing \cite{sereno2}, gravitational lensing by massless braneworld black holes \cite{eiroa1}, strong deflection lensing by charged black holes in scalar-tensor gravity \cite{eiroa2}, and the references therein.

Recently, Gibbons and Werner, introduced a new geometrical method for calculating the light deflection \cite{gibbons1}. Using the optical metric and applying the Gauss-Bonnet theorem to the optical geometry they calculated the deflection angle for some static  and stationary spacetime metrics \cite{gibbons2}. In this context, we used this method to calculate the deflection of light by charged black holes with topological defects \cite{kimet}, and more recently the deflection angle for finite distance for Schwarzschild-de Sitter and Weyl conformal gravity was investigated. \cite{asahi}. Furthermore, Werner extended this method to stationary metrics and calculated the deflection angle for the Kerr black hole \cite{werner}. The deflection angle around a static cosmic string and a global monopole was studied in the Ref. \cite{birrola,gott,gibbons3}, however, much less effort has been devoted to studies of the deflection angle by a spinning cosmic string \cite{eugen}. Motivated by this fact, we aim to calculate the deflection angle by a spinning cosmic string in EC theory of gravity.

Topological defects like domain walls, cosmic strings and monopoles may have been produced by phase transitions involving spontaneous symmetry breaking in the early universe \cite{kibble}.  Although without any experimental evidence whatsoever, cosmic strings have been widely studied in the literature. There are some interesting effects associated with the presence of topological defects, in particular, the vacuum fluctuations \cite{konkowski,matsas}, gravitational lensing of cosmic string/global monopole \cite{vilenkin}, finite electrostatic self-force on an electric charged particle \cite{linet}, Landau quantization \cite{muniz}, and many others.  Line defects containing torsion, like dislocations, appear within Einstein-Cartan gravitation theory \cite{hehl} in Riemann-Cartan geometry. Furthermore, it was shown that cosmic strings can be represented by disclinations in the spacetime and the helical structure of the spinning cosmic string is associated with the timelike and spacelike dislocations with torsion effects in the EC theory was investigated here \cite{gal,ozdemir}. In Ref. \cite{dias} the effects of torsion on the electromagnetic field is studied, in the Ref. \cite{nikodem} torsion as alternative to cosmic inflation is investigated. Therefore, in this context, it's natural to see if torsion effects the light deflection by a spinning cosmic string.

The remainder of the paper is organized as follows. In Section 2, we briefly review the Finsler geometry and then we introduce the String-Randers optical metric and discuss some of its basic features. In Section 3, we calculate the corresponding Gaussian optical curvature for a spinning cosmic string and then using the Gauss-Bonnet theorem we calculate the leading terms of the deflection angle in the weak limit approximation. In Section 4, we comment on our results.  

\section{String-Randers optical metric}
In order to deal with the corresponding optical geometry for some stationary spacetime metrics, one needs to replace the Riemannian geometry with the Finsler geometry. In particular, the Finsler metric on a smooth manifold $M$, is defined as a smooth, nonnegative function $F (x, X)$, with the Hessian given by \cite{werner,bao}
\begin{equation}
g_{ij}(x,X)=\frac{1}{2}\frac{\partial^{2}F^{2}(x,X)}{\partial X^{i}\partial X^{j}},\label{1}
\end{equation}
where  $F^ {2} (x, X) =g_ {ij} (x, X) X^ {i} X^ {j} $. Note that the Randers metric is a special Finsler metric, given by the metric 
\begin{equation}
F(x, X)=\sqrt{a_{ij}(x)X^{i}X^{j}}+b_{i}(x)X^{i},\label{2}
\end{equation}
where $a_{ij}$ denotes a Riemannian metric and $b_{i}$ a one-form satisfying the condition $a^{ij}b_{i}b_{j}<1$.
Furthermore, we can make use of the Hessian \eqref{1}, and construct the Christoffel symbols which are given by
\begin{equation}
\Gamma^{i}_{jk}(x,X)=\frac{1}{2}g^{il}(x,v)\left(\frac{\partial g_{lj}(x,X)}{\partial x^{k}}+\frac{\partial g_{lk}(x,X)}{\partial x^{j}}-\frac{\partial g_{jk}(x,X)}{\partial x^{l}}\right),\label{3}
\end{equation}
where $g^{ij}$ is the inverse of \eqref{1} and $v_{i}=g_{ij}(x,X)X^{j}$. The  metric of an infinitely long spinning cosmic string with torsion effects in cylindrical coordinates reads \cite{gal,ozdemir}
\begin{equation}
\mathrm{d}s^{2}=-(\mathrm{d}t+a\, \mathrm{d}\varphi)^{2}+\mathrm{d}\rho^{2}+\alpha^{2}\rho^{2}\mathrm{d}\varphi^{2}+\left(\mathrm{d}z+\beta \mathrm{d}\varphi \right)^{2}\label{4}
\end{equation}
where $a=4GJ^{t}$ is the rotational parameter of the string, which has unit of distance, $\beta=4GJ^{z}$ is analogous to Burgers vector and is related to the torsion. The cosmic string parameter is defined in terms of string energy density $\mu$  and Newton's constant $G$ as $\alpha=(1-4G\mu)$ and $0<\alpha\leq 1$ (From now on, we will always assume $G=1$). The constants $J^{t}$ and $J^{z}$ are the intrinsic spin of the string and dislocations of the space respectively. Using the  transformations $z=r \cos\theta$ and $\rho=r\sin\theta$, we can write the above metric in spherical coordinates.  Without loss of generality  by setting $\theta=\pi/2$, we can eliminate the cross-terms and study the equatorial plane, in this way the metric \eqref{4} reads
\begin{equation}
\mathrm{d}s^{2}=-\left(\mathrm{d}t+a\,\mathrm{d}\varphi\right)^{2}+\mathrm{d}r^{2}+\left(\alpha^{2}r^{2}+\beta^{2}\right) \mathrm{d}\varphi^{2}.\label{5}
\end{equation}

Note that this stationary metric is locally flat, however asymptotically conical due the presence of a cosmic string with energy density $\mu=10^{-5}$. The cosmic string is extended along the $\theta=0$ and $\theta=\pi$, say, in the $z$-axes. To see how the Randers metric \eqref{2},  arises from the spinning cosmic string metric \eqref{4}, let $\mathrm{d}t=F(x,\mathrm{d}x)$, and solve the last metric for null geodesics with $\mathrm{d}s^{2}=0$, if follows that 
\begin{eqnarray}\notag
a_{ij}\,\mathrm{d}x^{i}\,\mathrm{d}x^{j}&=&\mathrm{d}r^{2}+\left(r^{2}\,\alpha^{2}+\beta^{2}\right) \,\mathrm{d}\varphi^{2},\\
b_{i}\,\mathrm{d}x^{i}&=&-a\,\mathrm{d}\varphi.\label{6}
\end{eqnarray}

As we know, Fermat's principle of general relativity states that the light-rays are selected by stationary arrival time at the observer
\begin{equation}
0=\delta\,\int\limits_{\gamma}\mathrm{d}t=\delta\,\int\limits_{\gamma}F(x, \dot{x})\mathrm{d}t,\label{7}
\end{equation}
where $\gamma_{F}$ sadisfies the geodesic equation. Therefore the corresponding optical plane $(r,\varphi)$ describing the metric \eqref{5} is represented by the following String-Randers metric
\begin{equation}
F\left(r, \varphi, \frac{\mathrm{d}r}{\mathrm{d}t}, \frac{\mathrm{d}\varphi}{\mathrm{d}t}\right)=\sqrt{\left( \frac{\mathrm{d}r}{\mathrm{d}t}\right)^{2}+\left(r^{2}\alpha^{2}+\beta^{2}\right)\left( \frac{\mathrm{d}\varphi}{\mathrm{d}t}\right)^{2}}-a\, \frac{\mathrm{d}\varphi}{\mathrm{d}t}.\label{8}
\end{equation}

Next, one can now apply the Nazim's method to construct a Riemannian manifold $(M,\bar{g})$, osculating the Randars manifold $ (M, F) $. One can choose a smooth vector field $\bar{X}$ over $M$, that contains the tangent vectors along the geodesic $\gamma_{F}$, and hence $\bar{X}(\gamma_{F})=\dot{x}$. Then the Hessian \eqref{1} reads
\begin{equation}
\bar{g}_{ij}=g_{ij}(x,\bar{X}(x)),\label{9}
\end{equation}
with the corresponding Levi-Civita connection $\bar{\Gamma}^{i}_{jk}$. Without goining into details, by following the arguments presented here \cite{werner}, one can show that the geodesic $\gamma_{F}$ of $(M, F)$, is also a geodesic $\gamma_{\bar{g}}$ of $(M,\bar{g})$, which is quite a remarkable result. In other words, one can start from the optical geometry (in our case String-Randers metric), and then use the corresponding osculating Riemannian manifold to compute the deflection angle of light rays in the equatorial plane. By taking the line $r(\varphi)=b/\sin\varphi $ as zeroth approximation of the deflected angle, and taking into account only the leading terms of the vector field $\bar{X}=(\bar{X}^{r}, \bar{X}^{\varphi})(r, \varphi)$ near the boundary one can make the following choose: 
\begin{equation}
\bar{X}^{r}=\frac{\mathrm{d}r}{\mathrm{d}t}=-\cos\varphi+\mathcal{O}(a), \hspace{1cm}\bar{X}^{\varphi}=\frac{\mathrm{d}\varphi}{\mathrm{d}t}=\frac{\sin^{2}\varphi}{b}+\mathcal{O}(a).\label{10}
\end{equation}

\section{Optical curvature and the deflection angle}

The famous Gauss-Bonnet theorem relates the intrinsic geometry of the spacetime with its topology. For reasons presented in the previous section, we can now apply the Gauss-Bonnet theorem to the osculating Riemannian manifoldfor $(D_{R},\bar{g})$, for the region $D_ {R} $ in $M$, with boundary $\partial D_{R}=\gamma_{\bar{g}}\cup C_ {R}$ which states that \citep{gibbons1,gibbons2,werner}
\begin{equation}
\iint\limits_{D_{R}}K\,\mathrm{d}S+\oint\limits_{\partial D_{R}}\kappa\,\mathrm{d}t+\sum_{i}\epsilon_{i}=2\pi\chi(D_{R}),\label{11}
\end{equation}
where $K$ is the Gaussian curvature and $\kappa$ is the geodesic curvature, defined as
$ \kappa=\bar{g}\,(\nabla_{\dot{\gamma}}\dot{\gamma}, \ddot{\gamma})$, such that $\bar{g}(\dot{\gamma}, \dot{\gamma})=1$ in which $\ddot{\gamma}$ is the unit acceleration vector and $\epsilon_{i}$ is the corresponding exterior angle at the $i$\,-th vertex. As $R\to \infty$, both jump angles become $\pi/2$, hence $\theta_{O}+\theta_{S}\to \pi$. Since $D_ {R} $ is non-singular, than the Euler characteristic is $\chi(D_{R})=1$, finally we are left with
\begin{equation}
\iint\limits_{D_{R}}K\,\mathrm{d}S+\oint\limits_{\partial D_{R}}\kappa\,\mathrm{d}t=2\pi\chi(D_{R})-(\theta_{O}+\theta_{S})=\pi.\label{12}
\end{equation}

Since $\gamma_{\bar{g}}$ is geodesic, clearly than $\kappa(\gamma_{\bar{g}})=0$, therefore our goal first is  to calculate $\kappa(C_{R})\mathrm{d}t$ where $\kappa(C_{R})=|\nabla_{\dot{C}_{R}}\dot{C}_{R}|$, as $R\to \infty$. In this way for very large $R$ given by  $C_{R}:= r(\varphi)=R=const.$, the radial component of the geodesic curvature can be given by
\begin{equation}
\left(\nabla_{\dot{C}_{R}}\dot{C}_{R}\right)^{r}=\dot{C}_{R}^{\varphi}\left(\partial_{\varphi}\dot{C}_{R}^{r}\right)+\bar{\Gamma}^{r}_{\varphi \varphi}\left(\dot{C}_{R}^{\varphi}\right)^{2}.\label{13}
\end{equation}

The first term in the last equation vanishes, therefore we are left only with the second term. If we use the fact that $\bar{g}_{\varphi \varphi}\,\dot{C}_{R}^{\varphi}\dot{C}_{R}^{\varphi}=1$ and then compute the Levi-Civita connection $\bar{\Gamma}^{r}_{\varphi \varphi}$, it's not difficult to show that for very large $r(\varphi)=R=const$, the geodesic curvature gives, $\kappa(C_{R})\to R^{-1}$. On the other hand, for constant $R$, from the metric \eqref{8} it follows that $\mathrm{d}t=(\sqrt{\alpha^{2}R^{2}+\beta^{2}}-a)\mathrm{d}\,\varphi$, and hence 
\begin{equation}
\lim_{R\to\infty}\kappa(C_{R})\mathrm{d}t=\lim_{R\to \infty} \frac{1}{R}\left(\sqrt{\alpha^{2}R^{2}+\beta^{2}}-a\right)\mathrm{d}\,\varphi=\alpha \,\mathrm{d}\,\varphi.\label{14}
\end{equation}

This is not a surprising result since our spacetime is globally conical due to the presence of the cosmic string and therefore the corresponding optical metric is not asymptotically Euclidian. The last result can also be written as $\kappa(C_{R})\mathrm{d}t/\mathrm{d}\varphi=\alpha\neq 1$, which reduces to asymptotically Euclidean $\kappa(C_{R})\mathrm{d}t/\mathrm{d}\varphi=1$, only if one takes the limit $\alpha\to 1$. Now, if we use this result and go back to Eq. \eqref{11}, it follows that
\begin{eqnarray}
\iint\limits_{D_{R}}K\,\mathrm{d}S&+&\oint\limits_{C_{R}}\kappa\,\mathrm{d}t\overset{{R\to \infty}}{=}\iint\limits_{S_{\infty}}K\,\mathrm{d}S+\alpha \int\limits_{0}^{\pi+\hat{\alpha}}\mathrm{d}\varphi.\label{15}
\end{eqnarray}

In the weak deflection limit  we may assume that the light ray is given by $r(t)=b/\sin\varphi $ at zeroth order, using \eqref{12} and \eqref{15} for the deflection angle one finds
\begin{equation}
\hat{\alpha}=4\mu \pi-\frac{1}{\alpha}\int\limits_{0}^{\pi}\int\limits_{\frac{b}{\sin \varphi}}^{\infty}K\,\sqrt{\det \bar{g}}\,\mathrm{d}r\,\mathrm{d}\varphi,\label{16}
\end{equation}
where we have expressed the first term in terms of the cosmic string energy density $\mu$, using $\alpha=1-4\mu $, and have neglected higher order terms $\mu^{2}$. 

In order to solve the last integral we need to compute the quantity $K \mathrm{d}S$, where $\mathrm{d}S=\sqrt{\det \bar{g}}\,\mathrm{d}r\,\mathrm{d}\varphi$.  Therefore let us now first compute the metric components of the osculating Riemannian manifold by using the String-Randers optical metric \eqref{8}, and then by using the equations \eqref{1}, \eqref{9} and \eqref{10}, one finds  that:
\begin{align}
\bar{g}_{rr}&=1-\frac{a\,\left(\alpha^{2}\,r^{2}+\beta^{2}\right)\sin^{6}\varphi}{b^{3}\left(\cos^{2}\varphi +\frac{\left(\alpha^{2}\,r^{2}+\beta^{2}\right)\sin^{4}\varphi}{b^{2}}\right)^{3/2}}\\
\bar{g}_{\varphi \varphi}&=\alpha^{2}r^{2}+\beta^{2}-\frac{a\,\left(\alpha^{2}\,r^{2}+\beta^{2}\right)\sin^{2}\varphi \left(3\, \cos^{2}\varphi +\frac{2\left(\alpha^{2}\,r^{2}+\beta^{2}\right)\sin^{4}\varphi}{b^{2}}\right)}{b\left(\cos^{2}\varphi +\frac{\left(\alpha^{2}\,r^{2}+\beta^{2}\right)\sin^{4}\varphi}{b^{2}}\right)^{3/2}}+\mathcal{O}(a^{2})\\
\bar{g}_{r\varphi}&=\frac{a\,\cos^{3}\varphi}{\left(\cos^{2}\varphi +\frac{\left(\alpha^{2}\,r^{2}+\beta^{2}\right)\sin^{4}\varphi}{b^{2}}\right)^{3/2}}.\label{19}
\end{align}

Note that we have neglected higher order terms like $a^{2}$. On the other hand the determinant of this metric can be written as:
\begin{equation}
\det \bar{g}=\left(\alpha^{2}r^{2}+\beta^{2}\right)\left[1-\frac{3\,a\,\sin^{2}\varphi}{b\,\sqrt{\cos^{2}\varphi +\frac{\left(\alpha^{2}r^{2}+\beta^{2}\right)\sin^{4}\varphi}{b^{2}}}}\right]+\mathcal{O}(a^{2}).\label{20}
\end{equation}

The Gaussian optical curvature can be given by \cite{werner}
\begin{equation}
K=\frac{\bar{R}_{r\varphi r\varphi}}{\det \bar{g}}=\frac{1}{\sqrt{\det \bar{g}}}\left[\frac{\partial}{\partial \varphi}\left(\frac{\sqrt{\det \bar{g}}}{\bar{g}_{rr}}\,\bar{\Gamma}^{\varphi}_{rr}\right)-\frac{\partial}{\partial r}\left(\frac{\sqrt{\det \bar{g}}}{\bar{g}_{rr}}\,\bar{\Gamma}^{\varphi}_{r\varphi}\right)\right],
\label{21}
\end{equation}
in which by using the above metric components for the Levi-Civita connections we find:
\begin{eqnarray}
\bar{\Gamma}^{\varphi}_{rr}&=&\frac{3ab^{2}\cos\varphi \sin^{4}\varphi\left[\left(\alpha^{2}r^{2}+\beta^{2}\right)\sin^{3}\varphi+2\cos^{2}\varphi\left(\left(\alpha^{2}r^{2}+\beta^{2}\right)\sin\varphi-br\alpha^{2}\right)\right]}{2\left(\alpha^{2}r^{2}+\beta^{2}\right)\left(b^{2}\cos^{2}\varphi+\left(\alpha^{2}r^{2}+\beta^{2}\right)\sin^{4}\varphi\right)^{5/2}}+\mathcal{O}(a^{2}),\\
\bar{\Gamma}^{\varphi}_{r\varphi}&=&\frac{r\alpha^{2}}{\alpha^{2}r^{2}+\beta^{2}}+\frac{a r \alpha^{2}\sin^{6}\varphi \left[5b^{2}\cos^{2}\varphi+2\left(\alpha^{2}r^{2}+\beta^{2}\right)\sin^{4}\varphi\right]}{2\left(b^{2}\cos^{2}\varphi+\left(\alpha^{2}r^{2}+\beta^{2}\right)\sin^{4}\varphi\right)^{5/2}}+\mathcal{O}(a^{2}).\label{23}
\end{eqnarray}

Therefore one finds
\begin{equation}
K\mathrm{d}S=\left[-\frac{\alpha^{2}r^{2}}{\left(\alpha^{2}r^{2}+\beta^{2}\right)^{3/2}}-\frac{a}{4 \alpha^{3}\,r^{3}}f(r,\varphi)+\mathcal{O}(a^{2})\right]\mathrm{d}r \mathrm{d}\varphi.\label{24}
\end{equation}

As we can see, the Gaussian curvature of the optical metric $K$, is negative, and here comes the role of the global topology and the role of the Gauss-Bonnet theorem on the light deflection. The negative sign indicates that, locally, light rays diverge, therefore the light rays can converge only by considering the global topology of the spacetime. In the last equation  $f(r,\varphi)$ is a very long and complicated function of $r$ and $\varphi$ and the parameter $\alpha$, given by
\begin{equation}
f(r,\varphi)=\frac{\sin^{2}\varphi \left(A(r,\varphi)+B(r,\varphi)+C(r,\varphi)\right)}{\left(b^{2}\cos^{2}\varphi+r^{2}\alpha^{2}\sin^{4}\varphi\right)^{7/2}}.\label{25}
\end{equation}

In general, this function $f(r,\varphi)$ depends also on the parameter $\beta$. For the sake of simplicity, in the second term of the Eq. \eqref{24}, we will focus only the on the effects of the intrinsic spin of the string on the light deflection by setting $\beta\to 0$. Herein, the functions $A(r,\varphi)$, $B(r,\varphi)$ and $C(r,\varphi)$ are also some functions expressed in terms of $r$, $\varphi$ and the parameter $\alpha$, given by
\begin{eqnarray}\notag
A(r,\varphi)&=&r^{3}\alpha^{4}\sin\varphi\left[r^{3}\alpha^{2}\left(6b^{2}-r^{2}\alpha^{4}+r^{2}\alpha^{4}\cos(2\varphi)\right)\sin^{9}\varphi+12b^{4}\cos^{6}\varphi\left(4b-5r\sin\varphi\right)\right]\\\notag
B(r,\varphi)&=&-b^{2}r^{4}\alpha^{4}\cos^{2}\varphi\sin^{6}\varphi \left[24b^{2}-27r^{2}\alpha^{2}-r^{2}\alpha^{4}+r^{2}\alpha^{2}\left(27+\alpha^{2}\right)\cos(2\varphi)+36 b r \alpha^{2}\sin\varphi \right]\\\notag
C(r,\varphi)&=&b^{2}r^{3}\alpha^{4}\cos^{4}\varphi\sin^{3}\varphi \Big[36\, b\, r^{2}\alpha^{2}\cos(2\varphi)+r\left(45 r^{2}\alpha^{2}+b^{2}\left(34\alpha^{2}-66\right)\right)\sin\varphi+\\
&&3\left(8b^{3}-12 b r^{2}\alpha^{2}-5 r^{3}\alpha^{2}\sin(3\varphi) \right)\Big].\label{26}
\end{eqnarray}

In what follows we will make use of the last three equations and compute the deflection angle. Substituting the result \eqref{24} into the Eq. \eqref{16}, the deflection angle reads
\begin{equation}
\hat{\alpha}\approx 4\pi \mu+\frac{1}{\alpha}\int\limits_{0}^{\pi}\int\limits_{\frac{b}{\sin \varphi}}^{\infty}\left(\frac{\alpha^{2} \,\beta^{2} }{\left(\alpha^{2}r^{2}+\beta^{2}\right)^{3/2}}+\frac{a\,f(r,\varphi) }{4\,r^{3}\,\alpha^{3}}\right)\mathrm{d}r\mathrm{d}\varphi.
\label{27}
\end{equation}

The first integral reduces to complete elliptic integral of the first kind, after some approximations we find
\begin{equation}
\int\limits_{0}^{\pi}\int\limits_{\frac{b}{\sin \varphi}}^{\infty}\frac{\alpha^{2} \,\beta^{2} }{\left(\alpha^{2}r^{2}+\beta^{2}\right)^{3/2}}\mathrm{d}r\mathrm{d}\varphi=\frac{\beta^{2} \pi}{4 \,b^{2} \alpha^{2}},
\label{28}
\end{equation}
where $b$ is the impact parameter. On the other hand, after some long calculations, for the second integral we find the following result (see Appendix A)
\begin{equation}
\int\limits_{0}^{\pi}\int\limits_{\frac{b}{\sin \varphi}}^{\infty}\frac{a\,f(r,\varphi) }{4\,r^{3}\alpha^{3}}\mathrm{d}r\mathrm{d}\varphi=\frac{3\, a\, \pi}{8 \,b}\left(\frac{1}{\alpha}-\alpha\right).\label{29}
\end{equation}

Now using these results and going back to Eq. \eqref{27} for the total deflection angle we find
\begin{equation}
\hat{\alpha}\simeq 4\mu\pi+\frac{\beta^{2} \pi}{4 b^{2} \alpha^{3}}\pm \frac{3\, a\, \pi }{8 \,b}\left(1-\frac{1}{\alpha^{2}}\right), \label{30}
\end{equation}
in which the positive (negative) sign is for a retrograde (prograde) light ray, respectively. The first term refers to the deflection angle by a static cosmic string which is independent of the impact parameter $b$. Since $\mu=10^{-5}$, this angle is of the order of $3$ arcsec, which is small effect but within the observable range.  The second term is a direct consequence of the EC theory and is related to torsion. Finally, the third term is related to the intrinsic spin effects on the light deflection.

If now one assumes that, the light ray is propagating from the source $S$, to an observer $O$, such that both $S$ and $O$ lie on the same surface and are perfectly aligned \cite{vilenkin}, then, we can rewrite the deflection angle in terms of the mass density of the cosmic string and  $d$ and $l$, as follows 
\begin{equation}
\hat{\alpha}\simeq 4\mu\pi l\,\left(l+d\right)^{-1}+\frac{\beta^{2} \pi}{4 \,b^{2}}+\frac{3 \pi \beta^{2}\mu}{b^{2}} \pm\frac{3\pi a \mu }{b},\label{31}
\end{equation}
where $d$ and $l$ are the corresponding distances from the cosmic string to the observer $O$, and to the source $S$, respectively. From Eq. \eqref{31}, it is clear that, the effects of the intrinsic spin and torsion on the light diflection are negligible compared with the deflection angle of the static cosmic string. 

In the special case if $|J^{z}|=0$, then this metric represents a spinning cosmic string with no cosmic dislocations
\begin{equation}
\mathrm{d}s^{2}=-(\mathrm{d}t+a\, \mathrm{d}\varphi)^{2}+\mathrm{d}\rho^{2}+\alpha^{2}\rho^{2}\mathrm{d}\varphi^{2}+\mathrm{d}z^{2},
\label{rn}
\end{equation}
setting $a=1$, and $b=10$ pc, the deflection angle is extremely small $ \hat{\alpha}_{spin}\approx 10^{-23}$ rad.
On the other hand if we choose $|J^{t}|=0$, then, this metric represents a cosmic dislocations given by the following  metric
\begin{equation}
\mathrm{d}s^{2}=-\mathrm{d}t^{2}+\mathrm{d}\rho^{2}+\alpha^{2}\rho^{2}\mathrm{d}\varphi^{2}+\left(\mathrm{d}z^{2}+\beta \mathrm{d}\varphi\right)^{2}.
\label{rn}
\end{equation}

The torsion effects are even smaller, in fact, if we choose $\beta=1$, then $ \hat{\alpha}_{torsion}\approx 10^{-34}$rad, therefore from the practical point of view impossible to be detect experimentally. In the special case if $\alpha\to 1$, $\beta \to 0$ and $a\to0$, we recover the Minkowski spacetime, $\hat{\alpha}_{flat}=0$, as expected. However, will be interesting to see the torsion effects on light deflection if we apply this method to the Kerr black hole or some other black hole configuration.

\bigskip

\section{Conclusion}
In summery, we have calculated the leadin termes of the deflection angle in the weak limit approximation caused by a spinning cosmic string within the Einstein-Cartan theory of gravity. For this reasen we have introduced the String-Randers optical metric and applied the Gauss-Bonnet theorem to this optical metric.  The results shows that, the effects of the intrinsic spin of the string and torsion on light deflection are negligible compared with the deflection angle of the static cosmic string, nevertheless they are present in the final result for the total deflection angle. 

\section{Appendix A}

In order to solve the integral \eqref{29}, we can first integrate with respect to $r$ and find the following integral

\begin{equation}
I=\int\limits_{0}^{\pi}\int\limits_{\frac{b}{\sin \varphi}}^{\infty}\frac{a\,f(r,\varphi) }{4\,r^{3}\alpha^{3}}\mathrm{d}r\mathrm{d}\varphi=\int_{0}^{\pi}\Big[\left.\frac{a \alpha \sin^{2}\varphi \Big(M(r,\varphi)+N(r,\varphi)+Q(r,\varphi)\Big)}{b \sqrt{2}\Big(4b^{2}+3 r^{2}\alpha^{2}+4(b^{2}-r^{2}\alpha^{2})\cos(2\varphi)+r^{2}\alpha^{2}\cos(4\varphi)\Big)^{5/2}} 
\Big]\right\vert_{\frac{b}{\sin\varphi}}^{\infty}\mathrm{d}\varphi, \label{a1}
\end{equation}
where 
\begin{eqnarray}\notag
M(r,\varphi)&=&\sec^{2}\varphi \Big(b+b \cos(2\varphi)+10 r \sin\varphi+2 r \sin(3\varphi)\Big)\Big(4 b^{2}+3 r^{2}\alpha^{2}+4 (b^{2}-r^{2}\alpha^{2})\cos(2\varphi)+r^{2}\alpha^{2}\cos(4\varphi)\Big)^{2}\\\notag
N(r,\varphi)&=&48 \alpha^{-2}b^{4}\cot^{2}\varphi \Big(19b-b\alpha^{2}+12 b \cos(2\varphi)+b(1+\alpha^{2})\cos(4\varphi)+14r \alpha^{2}\sin\varphi-3 r \alpha^{2}\sin(3 \varphi)-r\alpha^{2}\sin(5\varphi)\Big)\\\notag
Q(r,\varphi)&=&-2b^{2}\alpha^{-2}\csc^{2}\varphi \Big(4b^{2}+3r^{2}\alpha^{2}+4(b^{2}-r^{2}\alpha^{2})\cos(2\varphi)+r^{2}\alpha^{2}\cos(4\varphi)\Big)\times\\
&&\Big(42 b+b \alpha^{2}+36 b \cos(2\varphi)-b(\alpha^{2}-2)\cos(4\varphi)-14r\alpha^{2}\sin(\varphi)+3r\alpha^{2}\sin(3\varphi)+r\alpha^{2}\sin(5\varphi)\Big).
\end{eqnarray}

Now one can go back to the Eq. \eqref{a1} and take the limit of the integrand when $r\to\infty$, and show that in fact the following integral vanishes
\begin{equation}
\int_{0}^{\pi}\Big(\frac{a\csc\varphi \sec^{2}\varphi \sin^{2}\varphi\,(3+\cos(2\varphi))}{b}\Big)\mathrm{d}\varphi=0.
\end{equation}

Therefore we are left only with the contribution of the second term. After evaluating the integral \eqref{a1} when $r=b/\sin\varphi$, we find the following result
\begin{equation}
I=\frac{a}{2 b \alpha}\frac{(1+\alpha^{2})E(\pi,m^{2})-2\alpha^{2}F(\pi,m^{2})}{1-\alpha^{2}},
\end{equation}
where $m^{2}=1-\alpha^{2}$. Note that $E(\pi,m^{2})$ and $F(\pi,m^{2})$ represents the complete elliptic integrals of the second and first kind, respectively. We can expand these functions in Taylor's series around the point $m$ and find
\begin{equation}
E(\pi,m^{2})\approx \pi - \frac{\pi m^{2}}{4}-\mathcal{O}(m^{4}),\,\,\,\,F(\pi,m^{2})\approx \pi + \frac{\pi m^{2}}{4}+\mathcal{O}(m^{4}).
\end{equation}

Finally making use of the last two equations we find the following result for the integral \eqref{a1} given by
\begin{equation}
I=\frac{3 a \pi}{8 \,b}\left(\frac{1}{\alpha}-\alpha\right).
\end{equation}

\end{document}